\documentclass{cs19proc}

\editors{G.~A. Feiden}
\publisher{Zenodo}
\conference{The 19th Cambridge Workshop on Cool Stars, Stellar Systems, and the Sun}
\conferencedate{2016}

\title{The multipolar magnetic fields of accreting pre-main-sequence stars: B at the inner disk, B along the accretion flow, and B at the accretion shock}
\author{Scott G. Gregory,$^{1}$ 
            Jean-Fran{\c c}ois Donati,$^{2,3}$
            Gaitee A. J. Hussain$^{4}$}

\affiliation{$^{1}$ School of Physics and Astronomy, University of St Andrews, St Andrews, KY16 9SS, U. K. \\
	$^{2}$Universit{\'e} de Toulouse, UPS-OMP, IRAP, 14 av Belin, F-31400 Toulouse, France \\ 
	$^{3}$CNRS, IRAP / UMR 5277, 14 av Belin, F-31400 Toulouse, France \\
	$^{4}$European Southern Observatory, Karl-Schwarzschild-Str. 2, 85748, Garching bei M{\"u}nchen, Germany}

\shorttitle{PMS star multipolar \mathbf{B}-fields}
\shortauthors{S. G. Gregory, J.-F. Donati \& G. A. J. Hussain}

\abs{Zeeman-Doppler imaging studies have revealed the complexity of the large-scale magnetic fields of accreting pre-main-sequence stars.  All have multipolar magnetic fields with the octupole component being the dominant field mode for many of the stars studied thusfar.  Young accreting stars with fully convective interiors often feature simple axisymmetric magnetic fields with dipole components of order a kilo-Gauss (at least those of mass $\gtrsim0.5\,{\rm M}_\odot$), while those with substantially radiative interiors host more complex non-axisymmetric magnetic fields with dipole components of order a few 0.1 kilo-Gauss.  Here, via several simple examples, we demonstrate that i). in most cases, the dipole component alone can be used to estimate the disk truncation radius (but little else); ii) due the presence of higher order magnetic field components, the field strength in the accretion spots is far in excess of that expected if a pure dipole magnetic field is assumed. (Fields of $\sim$6$\,{\rm kG}$ have been measured in accretion spots.); iii) if such high field strengths are taken to be representative of the polar strength of a dipole magnetic field, the disk truncation radius would be overestimated.  The effects of multipolar magnetic fields must be considered in both models of accretion flow and of accretion shocks.}

\begin{document}

\maketitle

\section{Introduction: the magnetic topology of accreting pre-main-sequence stars}\label{intro}
Over the past decade, high-resolution optical spectropolarimeters have greatly enhanced our ability to study stellar magnetism across the Hertzsprung-Russell diagram.  Low-mass, accreting, pre-main-sequence (PMS) stars are of particular interest, as they reveal the history of the Sun at a time when the planets of the Solar System were forming.  

The first magnetic maps of an accreting PMS star, V2129~Oph, were published by \citet{don07}.  Constructed using the Zeeman-Doppler imaging technique, they revealed the long-suspected multipolar nature of PMS magnetism. The maps are constructed from a time series of circularly polarised spectra, and for accreting stars, by simultaneously considering the polarisation information contained in photospheric absorption lines and in accretion-related emission lines. Magnetic maps have now been published for the following accreting PMS stars, most at more than one epoch:  V2129~Oph, BP~Tau, V2247~Oph, AA~Tau, TW~Hya, V4046~Sgr AB, GQ~Lup, DN~Tau, CV~Cha, and CR~Cha \citep{don07,don08,don10b,don10a,don11a,don11b,don11c,don12,don13,hus09}.  All of them have multipolar magnetic fields. The majority of the magnetic maps were obtained as part of the multi-year Magnetic Protostars \& Planets (MaPP) large-observing program with the ESPaDOnS spectropolarimeter at the Canada-France-Hawaii telescope, and its twin instrument NARVAL at T{\'e}lescope Bernard Lyot.  The MaPP program spawned several additional, multi-wavelength, ground and space-based observations (e.g. \citealt{arg11,arg12,kas11,ale12}), as well as multiple theoretical / modelling papers (e.g. \citealt{gre08,gre10,gre11,jar08,lon11,rom11,joh14}).

\begin{figure}
	\centering
	\includegraphics[width=0.88\linewidth]{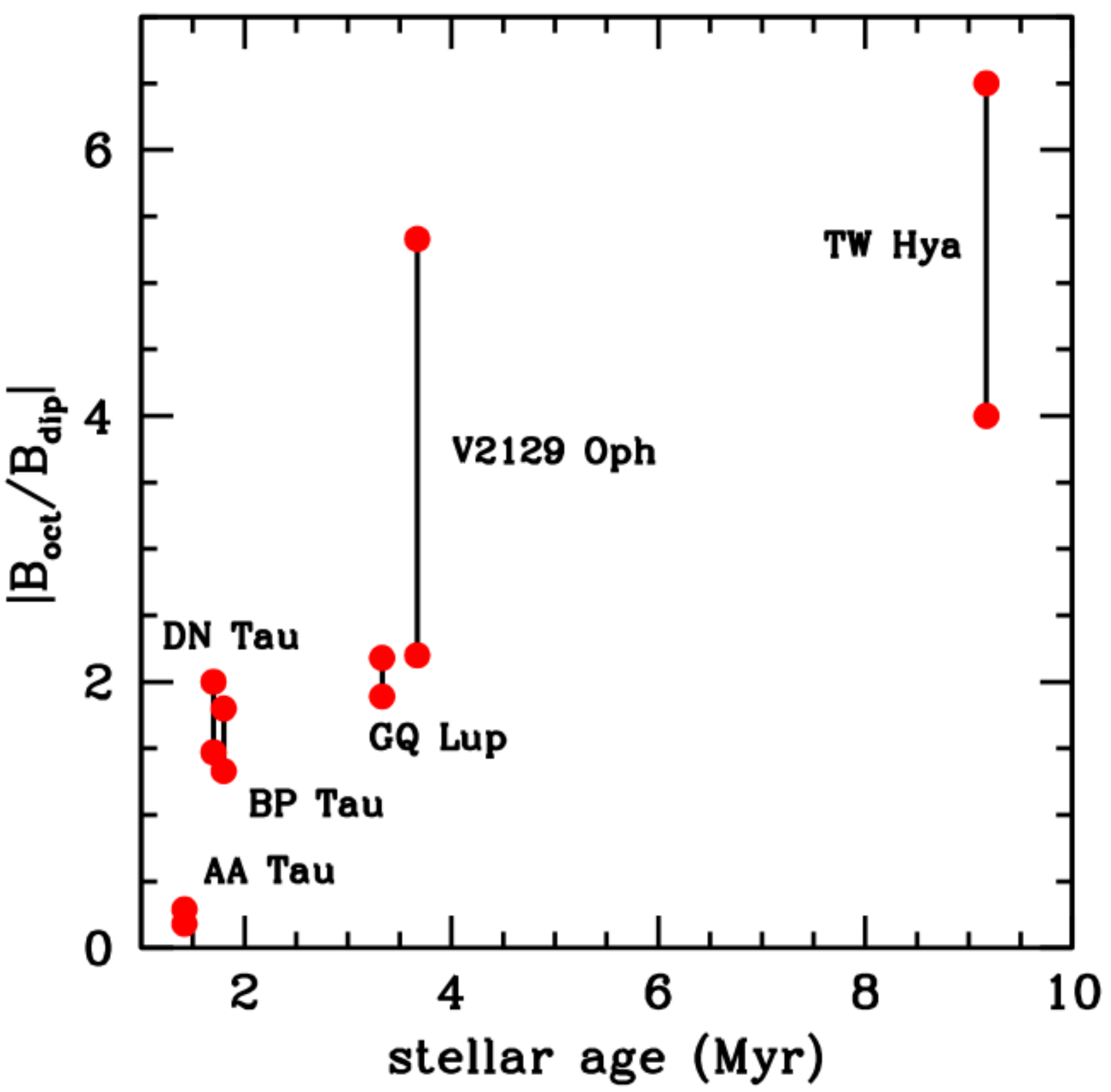}
	\caption{The magnitude of the ratio of the polar strength of the octupole to the dipole component, $|B_{\rm oct}/B_{\rm dip}|$, versus age for accreting PMS stars.  All stars shown are fully convective or have just developed radiative cores, based on their position in the Hertzsprung-Russell diagram.  The vertical bars connect stars observed at two epochs. Accreting PMS stars with published magnetic maps and more complex magnetic fields (5 stars), that are not well represented by a dipole-plus-octupole component, are not shown. Figure from \citet{gre14}.}
	\label{BoctBdip}
\end{figure}

As more magnetic topology information becomes available for accreting PMS stars it is becoming clearer that the internal structure of the star plays an important role in controlling the external, large-scale, magnetic field topology \citep{gre12,gre14}.  Accreting PMS stars, at least those more massive than $\sim$0.5$\,{\rm M}_\odot$, host strong axisymmetric large-scale magnetic fields while fully convective with the relative strength of the octupole to the dipole component increasing with age, see Figure \ref{BoctBdip}.  The large-scale magnetic field then becomes more complex and non-axisymmetric once the stellar interior becomes mostly radiative.  This stellar structure transition, and associated increase in magnetic field complexity, also has a signature in X-rays. The coronal X-ray emission decays once PMS stars have evolved onto Henyey tracks \citep{gre16}.  

Little is known about the magnetic topology of accreting PMS stars of mass $\lesssim$0.5$\,{\rm M}_\odot$.  However, based on the similarities between the magnetic topologies of main sequence M-dwarfs and of accreting PMS stars, see \citet{gre12}, it is expected that low-mass accreting PMS stars will show a variety of large-scale magnetic geometries, from simple and axisymmetric, to complex and non-axisymmetric.

Our goal in this conference proceedings is to highlight some of the ways in which multipolar magnetic fields influence magnetospheric accretion / the star-disk interaction and (hopefully) to clear up some lingering misconceptions that persist in the literature.  We do this using several straightforward, semi-analytic, back-of-the-envelope style calculations. In \S\ref{fields} we describe the field components for a stellar magnetic field consisting of a dipole plus an octupole component, an adequate first order approximation for the magnetosphere of many (but not all) accreting PMS stars (see \citealt{gre11} for extensive discussion). In \S\ref{disk} we demonstrate that, in most cases, the dipole component alone can be used to estimate the disk truncation radius, although little else in the star-disk system.  In \S\ref{column} we illustrate the strong departure of $B$ along the accretion column from that of a pure dipole.  In \S\ref{shock} we show that $B$ in the accretion shock can be multiple kilo-Gauss, even for accreting PMS stars with sub-kilo-Gauss dipole components, and that the disk truncation radius can be overestimated if $B$ at the accretion shock is (erroneously) assumed to be representative of a dipole large-scale magnetic field.  We conclude in \S\ref{conclusions}.


\section{Axisymmetric dipole-octupole magnetic fields}\label{fields}
Several accreting PMS stars (those plotted in Figure \ref{BoctBdip}) have large-scale magnetic fields that are well described by a tilted axisymmetric dipole component, plus a tilted axisymmetric octupole component, although in all cases higher order and non-axisymmetric multipole components are present too \citep{gre11}.  Some have large-scale magnetic fields where the dipole and octupole components are close to parallel, where the main positive pole of the dipole and of the octupole are in the same hemisphere. For other stars the main negative pole of one component is in same hemisphere as the main positive pole of the other, and the dipole and octupole moments are closer to an anti-parallel configuration.

For simplicity, and to make progress analytically, here we consider the straightforward example of an accreting PMS star with a disk in the midplane, hosting a magnetic field consisting of a parallel, and aligned, dipole component plus an octupole component. A detailed mathematical description of such magnetic fields, and the anti-parallel case, can be found in \citet{gre11} and Gregory {\it et al.} (in prep.).  

\begin{figure}
	\centering
	\includegraphics[width=0.88\linewidth]{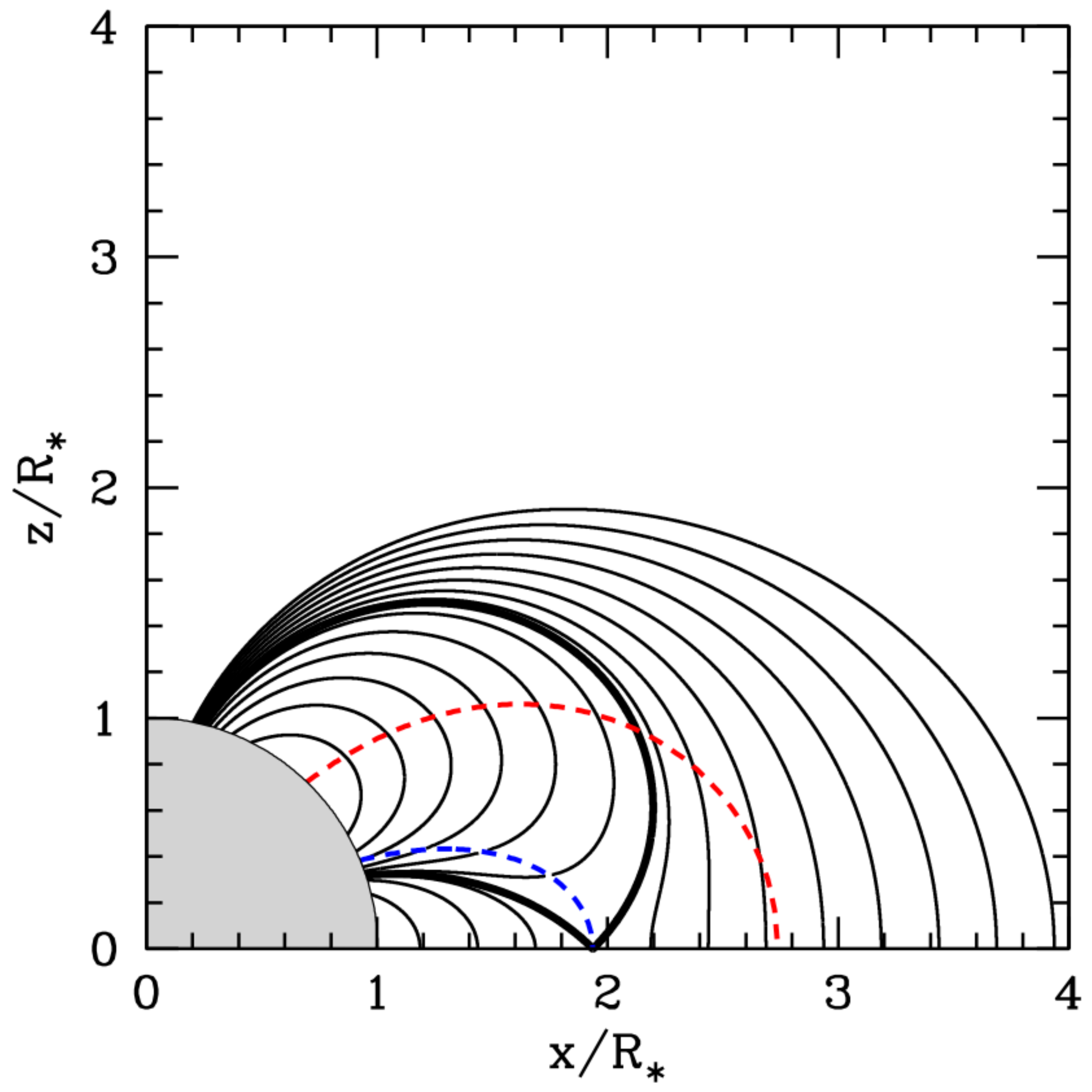}
	\caption{A star with a dipole-plus-octupole magnetic field where the dipole and octupole moments are parallel and aligned with the stellar rotation axis for $B_{\rm oct}/B_{\rm dip}=5$. $B_r=0$ along the dashed red line and $B_\theta=0$ along the dashed blue line. Field lines passing through the magnetic null point, see equation (\ref{rnull}), are highlighted in bold.}
	\label{fieldlineplot}
\end{figure}

Assuming the dipole and octupole magnetic moments are aligned with stellar rotation axis, then, in standard spherical polar coordinates $(r,\theta,\phi)$, the field components can be written as (see \citealt{gre10} for a derivation),
\begin{eqnarray}
B_r &=& B_{\rm dip} \left(\frac{R_\ast}{r} \right)^3\cos\theta \nonumber \\
       &+& \frac{1}{2}B_{\rm oct}\left(\frac{R_\ast}{r}\right)^5(5\cos^2\theta-3)\cos\theta, \label{equ_Br} \\
B_\theta &=& \frac{1}{2}B_{\rm dip} \left(\frac{R_\ast}{r}\right)^3\sin\theta \nonumber \label{equ_Bt} \\
        &+& \frac{3}{8} B_{\rm oct} \left(\frac{R_\ast}{r}\right)^5(5\cos^2\theta-1)\sin\theta, \\
B_\phi  &=& 0 \label{equ_Bp},
\end{eqnarray}
where $B_{\rm dip}$ and $B_{\rm oct}$ are the polar field strengths of the dipole and octupole field components respectively.  As the magnetic field being considered here is axisymmetric, $B_\phi=0$.  The field lines exterior to the star can be plotted by solving the differential equation,
\begin{equation}
\frac{B_r}{{\rm d}r} = \frac{B_\theta}{r{{\rm d}\theta}}.
\label{diffshape}
\end{equation}
An example for a star with $B_{\rm oct}/B_{\rm dip} = 5$ is shown in Figure \ref{fieldlineplot}. Note that the field topology depends on the ratio $B_{\rm oct}/B_{\rm dip}$ alone, although $B$ along the loops does depend on the values of $B_{\rm dip}$ and $B_{\rm oct}$.

For the case of parallel octupole and dipole magnetic moments considered here, it is clear that a magnetic null point (where all field components are zero) exists in the mid plane ($\theta=\pi/2$) at a radius of,
\begin{equation} 
\frac{r_{\rm null}}{R_\ast} = \left(\frac{3}{4}\frac{B_{\rm oct}}{B_{\rm dip}}\right)^{1/2}.
\label{rnull}
\end{equation}
$r_{\rm null}$ marks the transition point between field lines that connect the disk midplane to high latitudes on the stellar surface (for $r>r_{\rm null}$) and field lines that connect to lower latitudes (for $r<r_{\rm null}$), see \citet{gre11}.  In Figure \ref{fieldlineplot} the field lines that pass through $r_{\rm null}$ are highlighted in bold.  If the inner disk is truncated at $r<r_{\rm null}$ a portion, or all, of the accretion flow would impact the star at low latitudes. 


\section{B at the disk truncation radius}\label{disk}
The influence of multipolar magnetic fields (as constructed via field extrapolation from magnetic maps of accreting PMS stars) on the disk truncation radius $R_t$ has been considered by \citet{gre08} and \citet{joh14}.  Details of how dipole-plus-octupole magnetic fields affect $R_t$ can be found in \citet{ada12}.  Below, and in order to make progress analytically, we provide an overview of the results for dipole-plus-octupole magnetic fields.

In the equatorial plane, $\theta=\pi/2$ and from equations (\ref{equ_Br}-\ref{equ_Bp}), $B = |\mathbf{B}| = (B_r^2+B_\theta^2+B_\phi^2)^{1/2}$ reduces to,
\begin{equation}
B=\frac{1}{2}B_{\rm dip}\left(\frac{R_\ast}{r}\right)^3-\frac{3}{8}B_{\rm oct}\left(\frac{R_\ast}{r}\right)^5.
\label{Bmid}
\end{equation}
It is immediately obvious that the influence of the octupole component compared to that of the dipole diminishes rapidly with increasing distance from the star (and even more so for higher order magnetic field components not being considered in this simple example).  For typical disk truncation radii of $R_t \approx 5-10 R_\ast$ the relative contribution of the octupole compared to the dipole component to $B$ in the midplane [to $B$ in equation (\ref{Bmid})] is $(3/100)(B_{\rm oct}/B_{\rm dip})$ to $(3/400)(B_{\rm oct}/B_{\rm dip})$. Observed values of $|B_{\rm oct}/B_{\rm dip}|$ range from $\sim0.25-6$, see Figure \ref{BoctBdip}, with most $\lesssim2$.  With $B_{\rm oct}/B_{\rm dip}=2$ the contribution to B at the disk truncation radius from the octupole component is only 6\% that of the dipole component for $R_t=5R_\ast$, dropping to 1.5\% for $R_t=10R_\ast$.  Notice from equation (\ref{Bmid}) that B at the inner disk is less than it would be for a pure dipole, which will result in a smaller disk truncation radius for the dipole-plus-octupole magnetic fields.\footnote{If the dipole and octupole moments were anti-parallel then $R_t$ would be larger than for a pure dipole.}     

\begin{figure*}
	\centering
	\includegraphics[width=0.44\linewidth]{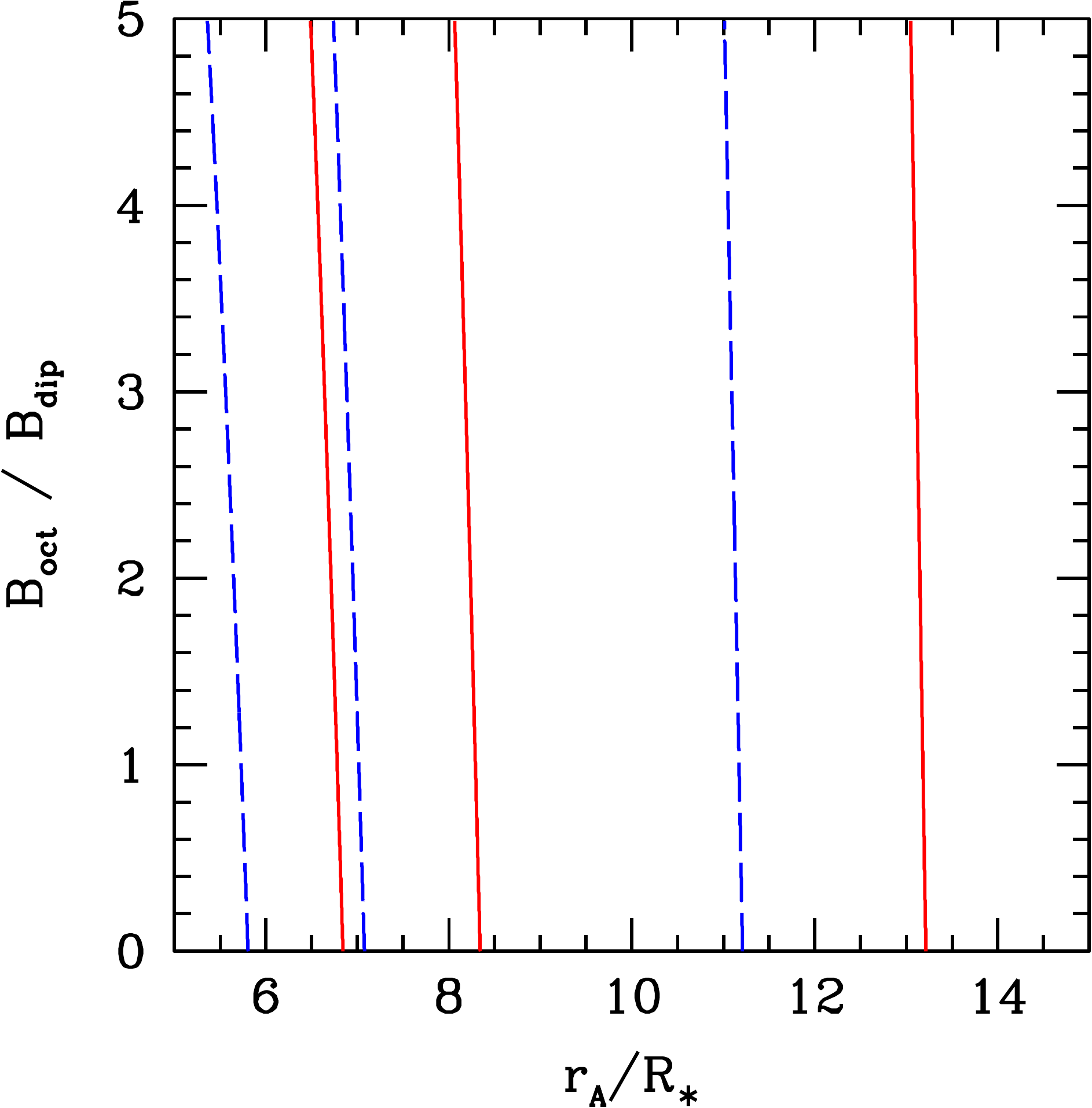}
	\includegraphics[width=0.44\linewidth]{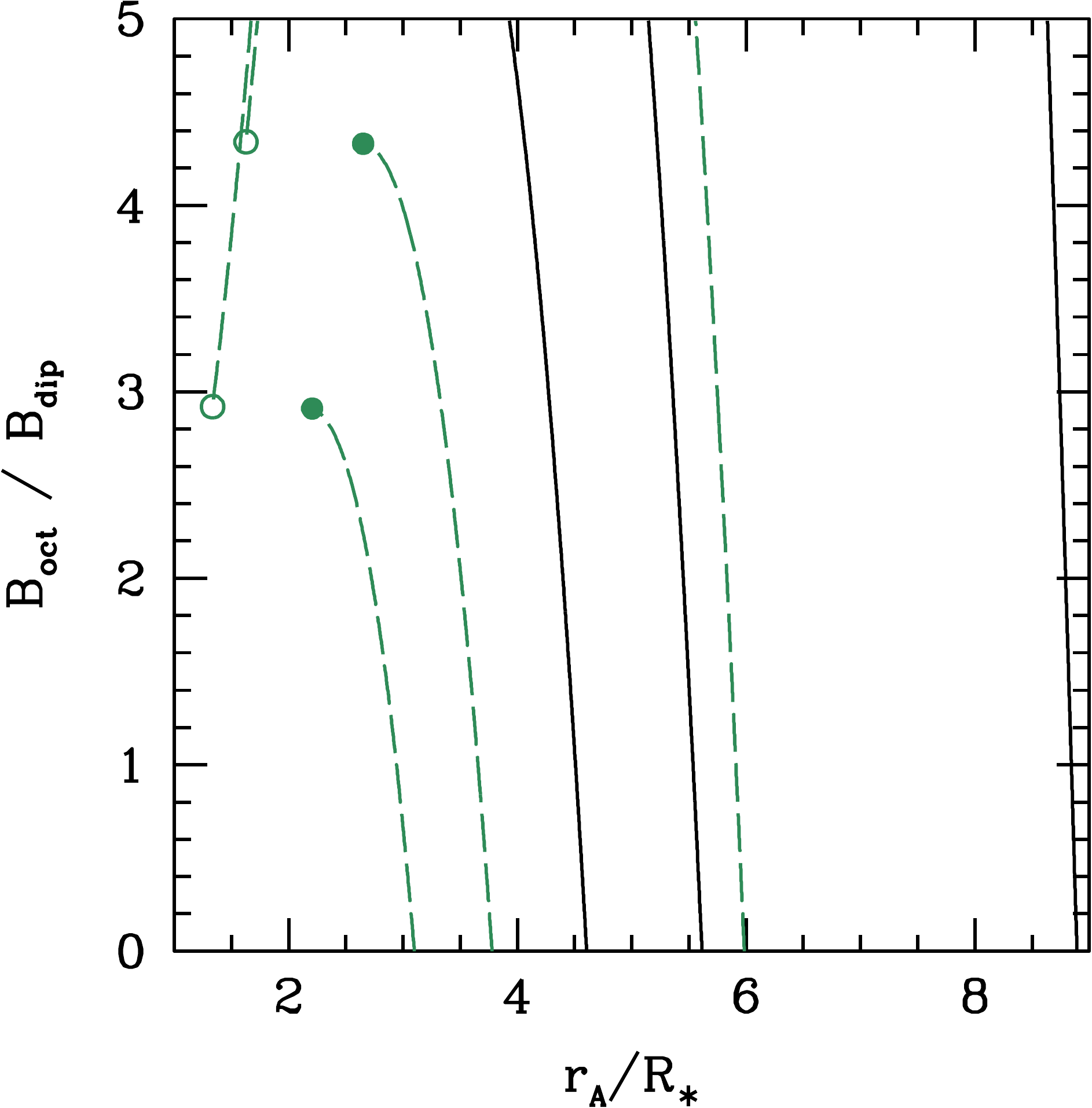}
	\caption{The variation of $r_A/R_\ast$ for different ratios of $B_{\rm oct}/B_{\rm dip}$ for $B_{\rm dip}=2, 1.5, 1$, and $0.5\,{\rm kG}$ (solid red, dashed blue, solid black \& dashed green lines respectively), with $B_{\rm oct}$ allowed to vary. Each set of coloured lines corresponds to a mass accretion rate of $\dot{M}=1{\rm e}$-$8, 5{\rm e}$-$9$, and $1{\rm e}$-$9\,{\rm M}_\odot{\rm yr}^{-1}$(leftmost line to rightmost line, respectively).  A stellar mass and radius of $0.7\,{\rm M}_\odot$ and $2\,{\rm R}_\odot$ have been assumed. Note the discontinuity in $r_A/R_\ast$, illustrated with filled/open circles, in two of the cases. The discontinuity arises as $r_A$ moves within $r_{\rm null}$, see equation (\ref{rnull}), and is due to the decrease in the magnetic pressure around the magnetic null point (see Figure \ref{Rt_pressure}) that exists in the midplane when the dipole and octupole moments are parallel. In most cases, $B_{\rm dip}$ alone can be used to calculate the disk truncation radius, with little change in $r_A/R_\ast$ with increasing $B_{\rm oct}$, with some exceptions (see text).}
	\label{Rtall}
\end{figure*} 

The location of the disk truncation radius not only depends on $B$ (and therefore on $B_{\rm dip}$ and $B_{\rm oct}$) but also on the mass accretion rate through the disk, $\dot{M}$.  This is not necessarily the $\dot{M}$ that ultimately accretes through columns on to the star, as a fraction may be diverted to outflows (e.g. \citealt{moh08}).  The disk is truncated at a fraction of the radius where the magnetic pressure provided by the stellar magnetosphere is balanced with the ram pressure of the bulk flow of material through the disk.  This is the Alfv{\'e}n radius, $r_A$, reduced by a numerical constant to account for the difference between spherical accretion and magnetospheric accretion (e.g. \citealt{kon91,joh14}).  For spherical infall at the free-fall speed, $v = (2GM_\ast/r)^{1/2}$, equating the magnetic energy density with the kinetic energy density, $B^2/(8\pi)=\rho v^2/2$, and using the equation of mass continuity $\dot{M}=4\pi r^2 \rho v$, gives at the Alfv{\'e}n radius,
\begin{equation}
B^2 = (2GM_\ast)^{1/2}\dot{M}r_A^{-5/2}.
\label{Bsq}
\end{equation}         
Equation (\ref{Bsq}) is evaluated in the midplane ($\theta=\pi/2$) with $B$ given by equation (\ref{Bmid}).  If the field was a pure dipole, $B_{\rm oct}=0$, then using equation (\ref{Bmid}), equation (\ref{Bsq}) reduces to the well known result,
\begin{equation}
R_{t,{\rm dip}} = cr_{A,{\rm dip}} = c \frac{\mu_{\rm dip}^{4/7}}{(2GM_\ast)^{1/7}\dot{M}^{2/7}},
\label{Rtdip}
\end{equation}    
where $\mu_{\rm dip} = B_{\rm dip}R_\ast^3/2$ is the dipole moment\footnote{Many literature sources use $\mu_{\rm dip}=B_\ast R_\ast^3$, where $B_\ast$ is the strength of the dipole at the stellar equator.  At the pole $B_{\rm dip} = 2B_\ast$ \citep{gre10}.}, $c$ is the constant ($<1$) that accounts for the difference between spherical infall and accretion along columns from the inner disk to the stellar surface,
and we have added ``dip'' subscripts to the radius terms to emphasis that the equation is valid for dipole magnetic fields.  The magnetohydrodynamic simulations of \citet{lon05} of the star-disk interaction with a dipole magnetic field suggest that $c = 1/2$, although this may not be applicable for multipolar magnetic fields.  Therefore, in the plots described below, we plot $r_A/R_\ast$ as the abscissa and remind readers that the true disk truncation radius is $R_t/R_\ast = cr_A/R_\ast$.  

\begin{figure*}
	\centering
	\includegraphics[width=0.44\linewidth]{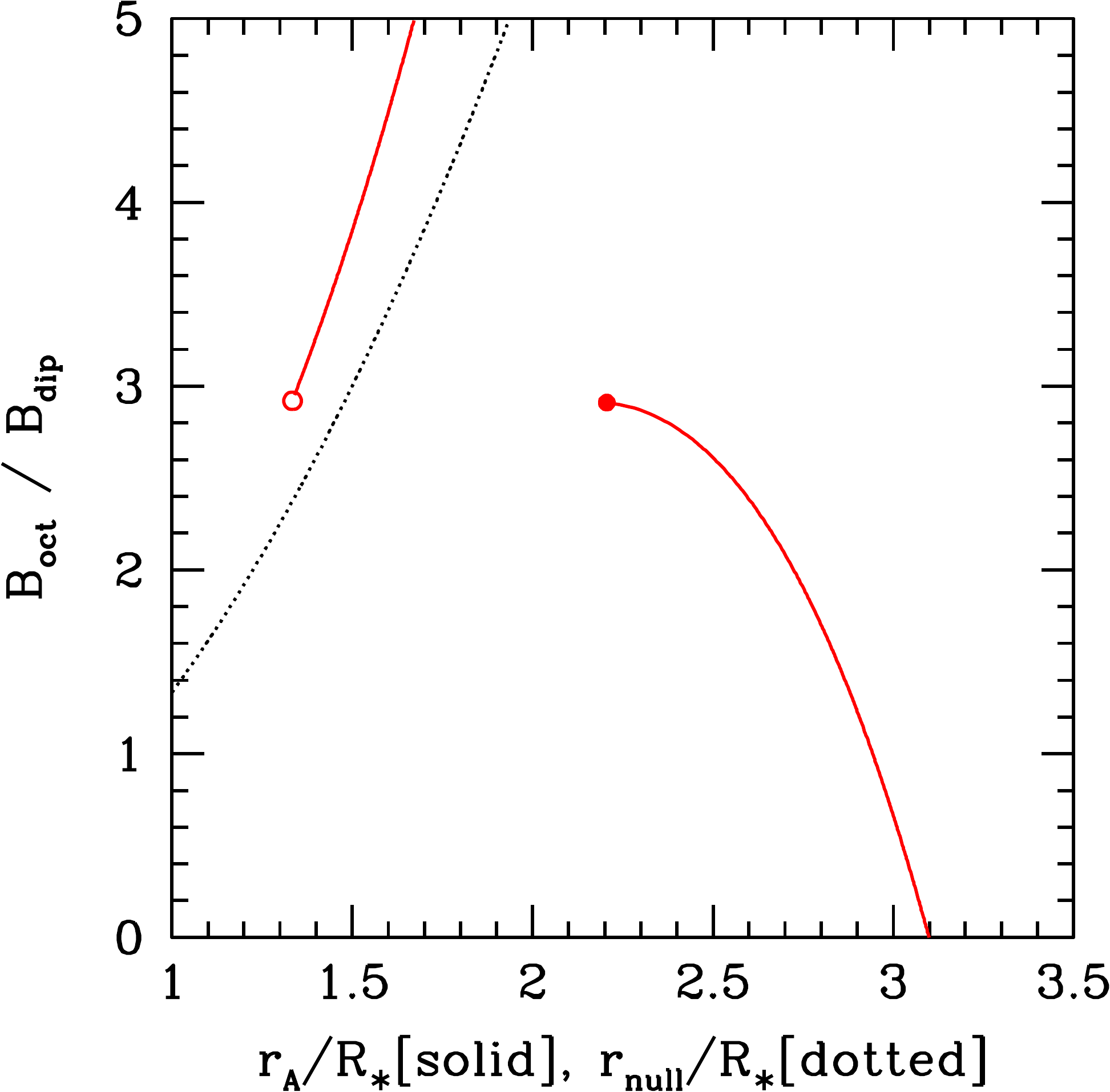}
	\includegraphics[width=0.44\linewidth]{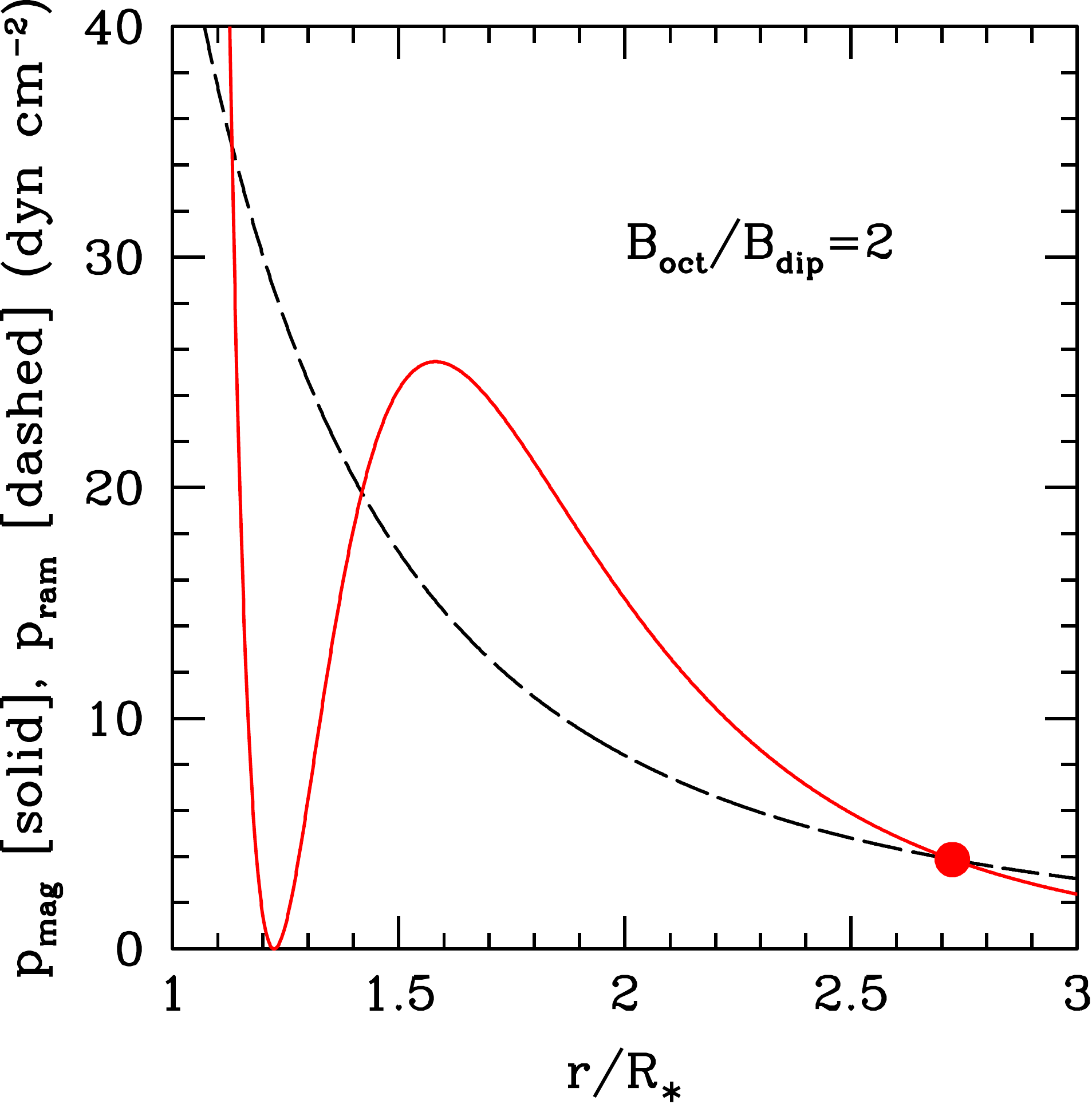}\\
	\includegraphics[width=0.44\linewidth]{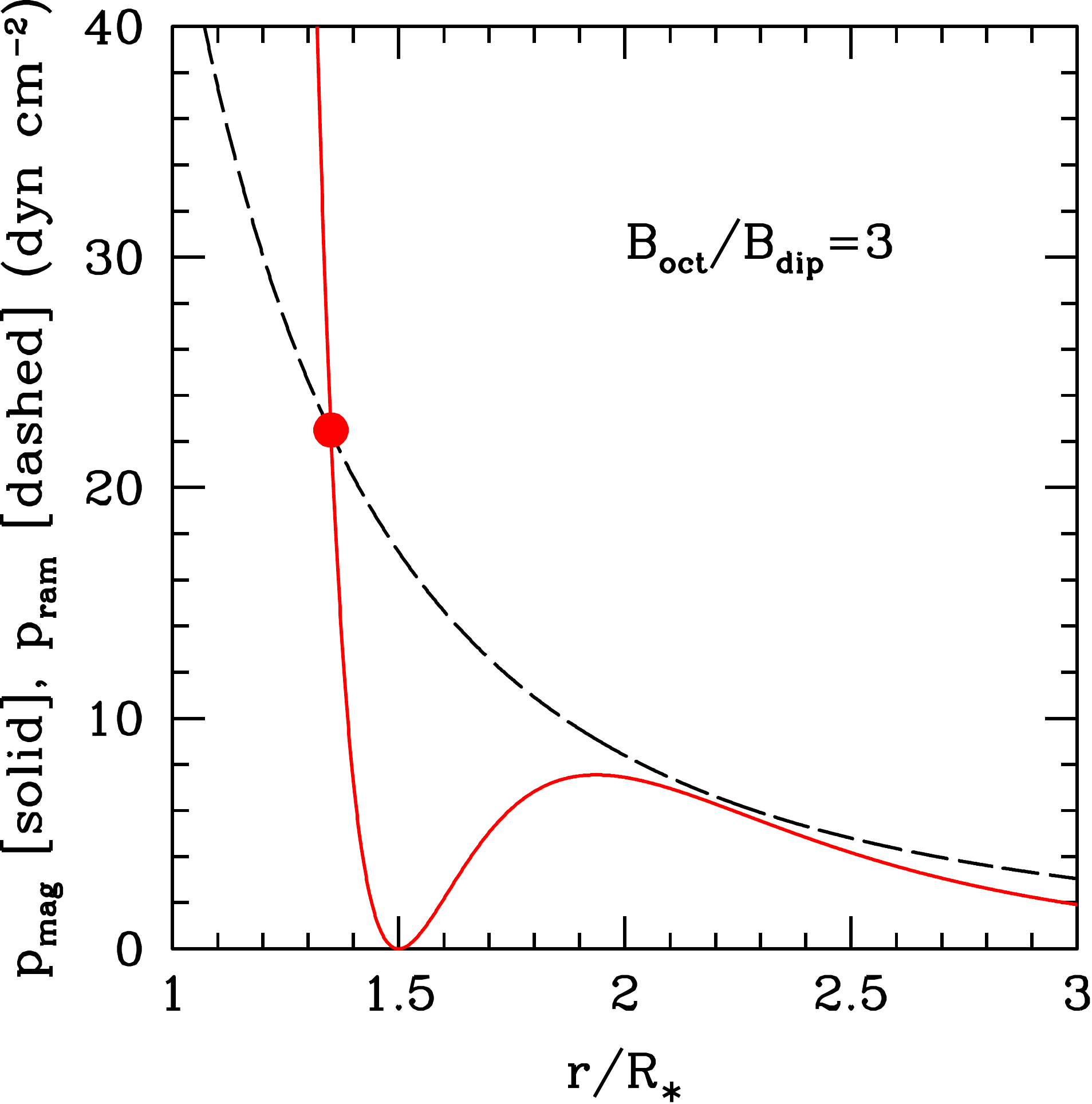}
	\includegraphics[width=0.44\linewidth]{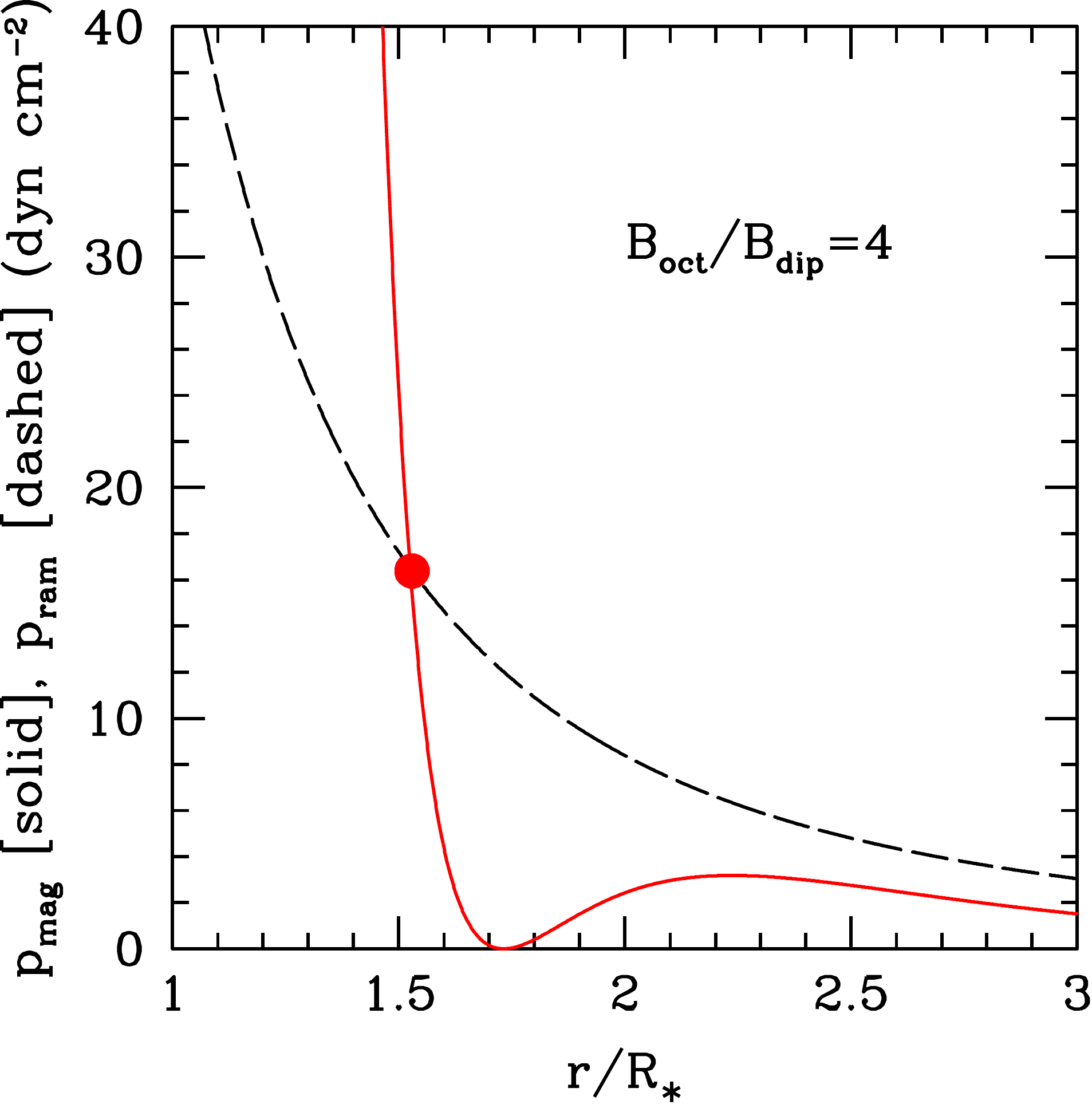}
	\caption{{\it Top left}: the variation of $r_A/R_\ast$ with $B_{\rm oct}/B_{\rm dip}$ calculated from equation (\ref{allt}) for $B_{\rm dip}=500\,{\rm G}$ (with $B_{\rm oct}$ allowed to vary), $\dot{M}=1{\rm e}$-$9\,{\rm M}_\odot{\rm yr}^{-1}$, $M_\ast=0.7\,{\rm M}_\odot$, and $R_\ast=2\,{\rm R}_\odot$. The dotted line shows the change in the magnetic null radius, equation (\ref{rnull}), with increasing $B_{\rm oct}/B_{\rm dip}$. The other three panels show the variation along the disk midplane of the magnetic pressure, $p_{\rm mag}$ (solid red line), and the ram pressure of the disk material, $p_{\rm ram}$ (dashed black line), calculated from the left- and right-hand-side of equation (\ref{bigequ}) respectively, for the indicated values of $B_{\rm oct}/B_{\rm dip}$.  The large red point shows $r_A/R_\ast$ in each case. In this example $r_A/R_\ast$ initially decreases, and then increases once again with increasing $B_{\rm oct}/B_{\rm dip}$, see the discussion in section \ref{disk}.} 
	\label{Rt_pressure}
\end{figure*}  

For the dipole-plus-octupole magnetic fields considered here, $r_A$ can again be calculated by equating the magnetic pressure and the ram pressure of the disk material in the midplane,
\begin{eqnarray}
\frac{1}{8\pi}\Bigg[\frac{1}{2}B_{\rm dip}\left(\frac{R_\ast}{r_A}\right)^3 &-& \frac{3}{8}B_{\rm oct}\left(\frac{R_\ast}{r_A}\right)^5\Bigg]^2 \nonumber \\ &=& \frac{1}{8\pi}(2GM_\ast)^{1/2}\dot{M}r_A^{-5/2},
\label{bigequ}
\end{eqnarray} 
which, using equation (\ref{Rtdip}), can be written as,
\begin{equation}
\frac{r_A}{R_\ast}\left[1-\frac{3}{4}\frac{B_{\rm oct}}{B_{\rm dip}}\left(\frac{R_\ast}{r_A}\right)^2\right]^{-4/7} = \frac{r_{A,{\rm dip}}}{R_\ast}.\label{allt}
\end{equation}
Equation (\ref{allt}) can be expanded to leading order, see \citet{ada12}, although it is straightforward to solve the full equation numerically for $r_A/R_\ast$, with the disk truncation radius then $R_t/R_\ast = cr_A/R_\ast$.   


In Figure \ref{Rtall} we demonstrate how the disk truncation radius changes as a function of $B_{\rm oct}/B_{\rm dip}$ for different mass accretion rates and various strengths of the dipole component.\footnote{For the accreting PMS stars with large-scale magnetic fields that are well described by a tilted dipole plus a titled octupole component, listed in section \ref{intro}, $B_{\rm dip}$ ranges from $\sim$0.3 to $\sim$1.9$\,{\rm kG}$ and $B_{\rm oct}$ from $\sim$0.5 to $\sim$2.8$\,{\rm kG}$, with values of $|B_{\rm oct}/B_{\rm dip}|$ as plotted in Figure \ref{BoctBdip}.}  A larger mass accretion rate and/or a weaker dipole component and/or a stronger octupole component (albeit to a lesser extent than the other quantities) corresponds to a smaller disk truncation radius.  It is also clear, that in most cases, the disk truncation radius is well approximated by the polar strength of the stellar dipole component alone (i.e. there is little variation in $r_A/R_\ast$, and therefore in $R_t/R_\ast$, with increasing $B_{\rm oct}/B_{\rm dip}$).  Exceptions to this are: (i) stars with very weak dipole components (or equivalently very strong higher order magnetic components); (ii) star-disk systems with large mass accretion rates; (iii) stars with highly tilted large-scale magnetospheres, where the field threading the disk midplane departs strongly from the vertical direction. All of these exceptions would allow the inner disk to push closer to the star, where the influence of higher order magnetic components is greater.  This can be seen in Figure \ref{Rtall}, where for parameters that result in smaller disk truncation radii, the change in $r_A$ with increasing $B_{\rm oct}/B_{\rm dip}$ is more significant.

\begin{figure*}
	\centering
	\includegraphics[width=0.44\linewidth]{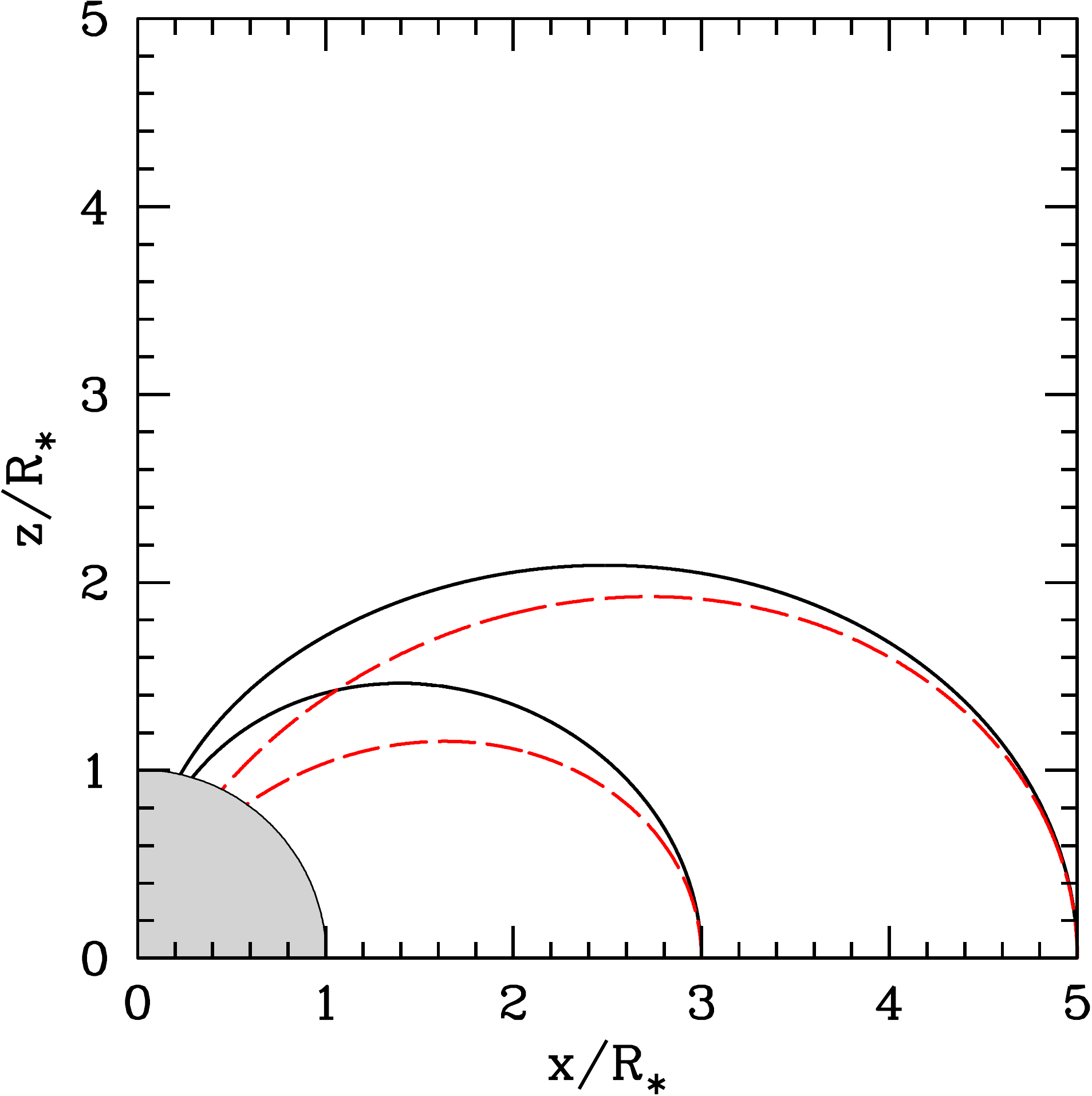}
	\includegraphics[width=0.44\linewidth]{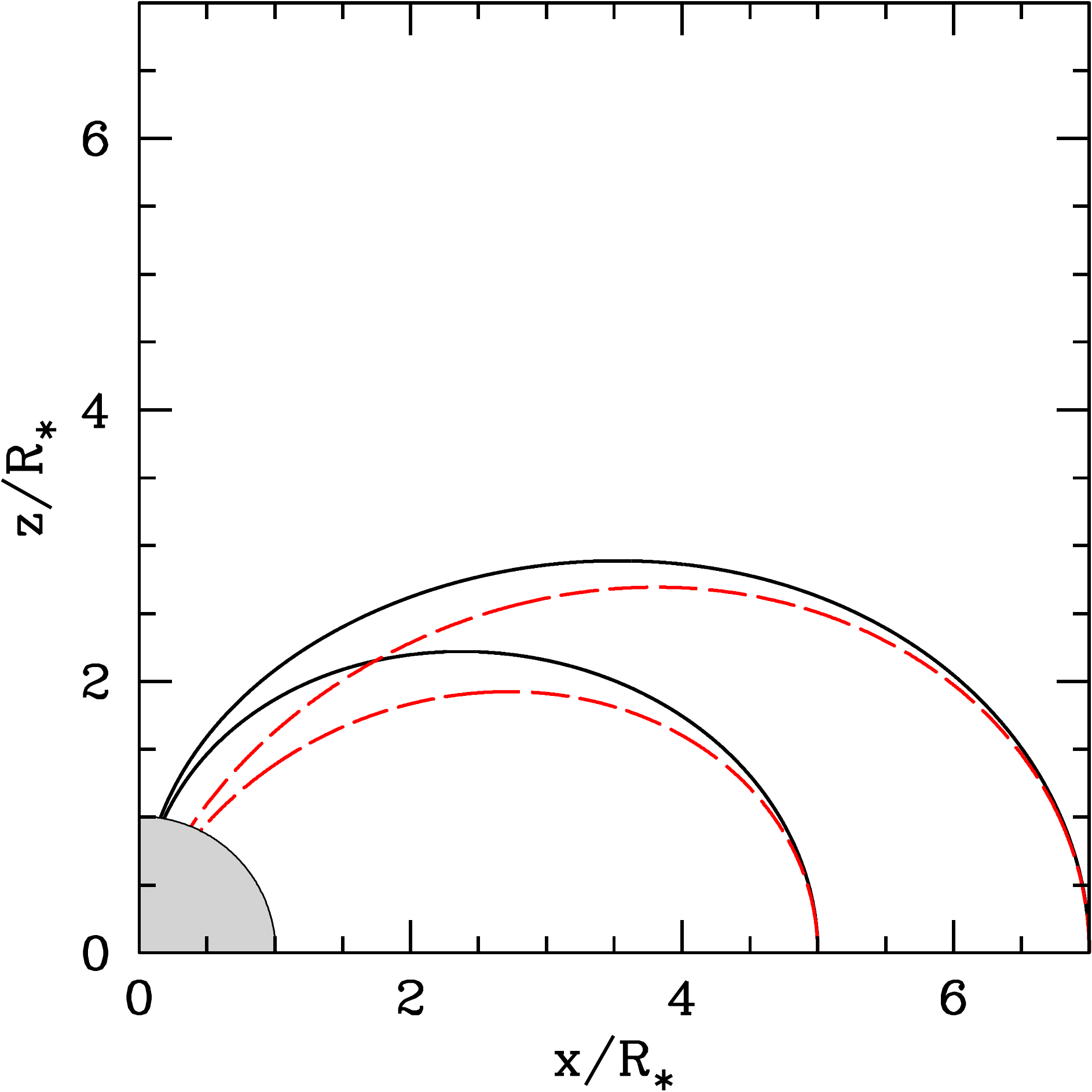}\\
	\includegraphics[width=0.44\linewidth]{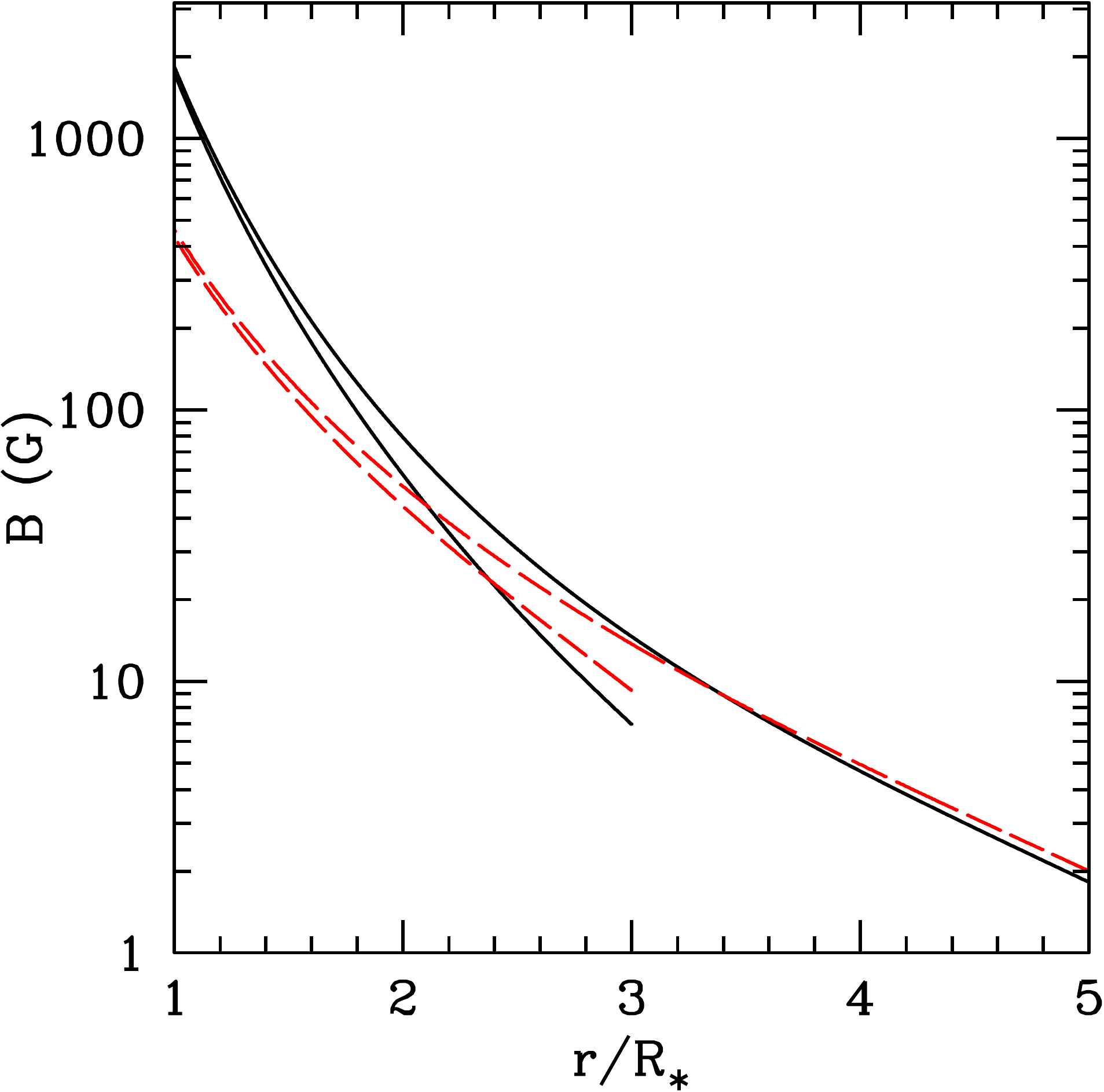}
	\includegraphics[width=0.44\linewidth]{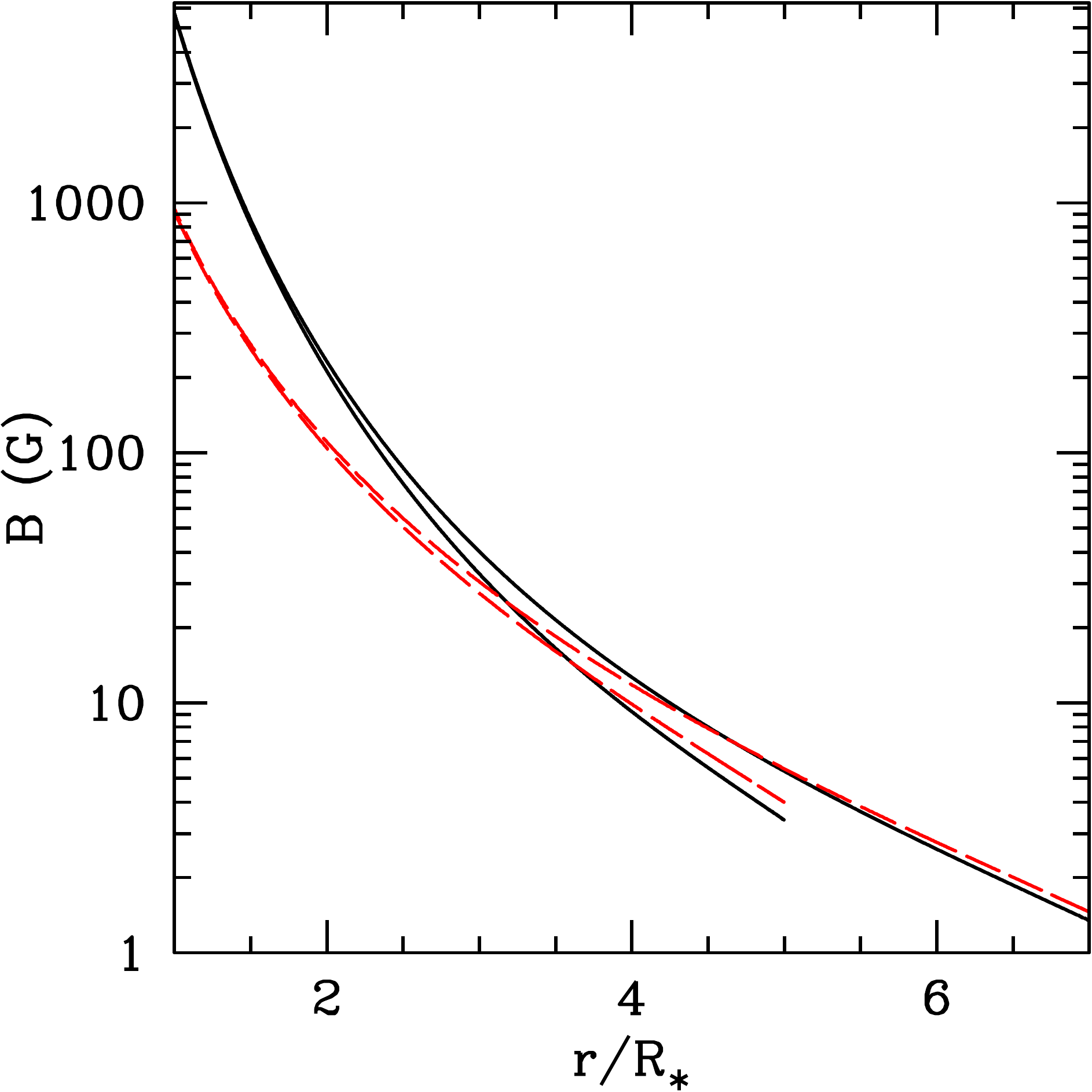}
	\caption{{\it Top row}: Some magnetic field lines for a star with parallel dipole and octupole moments [solid black lines; described by equation (\ref{dipoctshape})] with dipole magnetic loops for comparison (dashed red lines; plotted from the same radius in the disk midplane), which represent the accretion column for a typical set of PMS star parameters. The top left plot has field lines drawn from $r_m=3$ and $5R_\ast$ in the midplane to the stellar surface for $B_{\rm oct}/B_{\rm dip}$ = 3. The top right is for $r_m=5$ and $7R_\ast$ with $B_{\rm oct}/B_{\rm dip}=5$. {\it Bottom row}: $B$ along the same field lines sketched in the top row, with solid black lines (dashed red lines) again the dipole-plus-octupole (pure dipole) magnetic loops assuming $B_{\rm dip}=0.5\,{\rm kG}$ and $B_{\rm dip}=1\,{\rm kG}$ for the left and right plots respectively (which, given the $B_{\rm oct}/B_{\rm dip}$ ratios, is $B_{\rm oct}$=1.5\,{\rm kG} and $B_{\rm oct}=5\,{\rm kG}$ respectively). $B$ at the accretion shock can reach a few kilo-Gauss, even for stars with sub-kilo-Gauss dipole components.}
	\label{Bfieldlines}
\end{figure*}  

For certain parameters there is a discontinuity where the inner disk edge moves closer to the star, and then increases in radius again as $B_{\rm oct}/B_{\rm dip}$ is further increased.  A particular case is highlighted in Figure \ref{Rt_pressure}, and this behaviour can be understood as follows.  Starting at a large radius in the midplane, and moving towards the star, both the gas and magnetic pressures increase.  If the ratio $B_{\rm oct}/B_{\rm dip}$ and/or $\dot{M}$ is small, then the disk is truncated at a radius greater than the magnetic null point, see Figure \ref{Rt_pressure} top panels.  As we move further towards the star the magnetic pressure increases to a maximum\footnote{This maximum occurs at $r/R_\ast = [5B_{\rm oct}/(4B_{\rm dip})]^{1/2}$ and corresponds to the point in the midplane where the radius of curvature of the field lines is infinite (Gregory {\it et al.}, in prep.). Between this radius and $r_{\rm null}$ field lines are pinched towards the null point, see Figure \ref{fieldlineplot}.} before decaying to zero at $r_{\rm null}$. Once $r<r_{\rm null}$ the magnetic pressure increases again, rapidly, towards the stellar surface in the portion of the magnetosphere dominated by the octupole component. The minimum in the magnetic pressure means that as $B_{\rm oct}/B_{\rm dip}$ is increased further, see Figure \ref{Rt_pressure} bottom left panel, the disk truncation radius suddenly jumps from $r>r_{\rm null}$ to $r<r_{\rm null}$.  Further increasing $B_{\rm oct}/B_{\rm dip}$ then increases the disk truncation radius, see Figure \ref{Rt_pressure} bottom right panel, as the contribution to the magnetic pressure from the octupole component becomes increasingly significant.  This has the implication that for some stars a portion, or all, of the accretion flow will be channeled into low latitude hot spots; and that small variations in the polar strengths of the magnetic components and/or the mass accretion rate will allow the accretion flow to switch between high and low latitude hot spots (or a combination of both), altering the observed variability.


\section{B along the accretion column}\label{column}
In the previous section we demonstrated that the dipole component alone can, in many cases, provide an adequate estimate of the disk truncation radius. However,  the dipole component provides a poor approximation to the field strength along the accretion column, the shape of magnetic loops carrying accretion flow, and to the field strength at the accretion shock, as we demonstrate below. 

In order to calculate the field strength along loops carrying accreting gas, we must derive an equation that describes the path of the magnetic field lines from the disk to the stellar surface.  This is achieved by solving the differential equation (\ref{diffshape}). For a dipole [equations (\ref{equ_Br}) and (\ref{equ_Bt}) with $B_{\rm oct}$=0], and integrating from the midplane $(r=r_m,\theta=\pi/2)$ to a point $(r,\theta)$ along the magnetic loop, equation (\ref{diffshape}) yields the simple result, $r/R_\ast = (r_m/R_\ast)\sin^2\theta$.  For the parallel dipole-plus-octupole magnetic fields considered here, the equivalent field line equation is, 
\begin{eqnarray} 
&& \left[\left(\frac{r_m}{R_\ast}\right)^2 -\frac{1}{4}\frac{B_{\rm oct}}{B_{\rm dip}} \right]\left(\frac{r}{R_\ast}\right)^3-\left(\frac{r_m}{R_\ast}\right)^3\sin^2\theta\left(\frac{r}{R_\ast}\right)^2 \nonumber \\
 &-& \frac{1}{4}\left(\frac{r_m}{R_\ast}\right)^3\frac{B_{\rm oct}}{B_{\rm dip}}(5\cos^2\theta-1)\sin^2\theta=0,
 \label{dipoctshape}
\end{eqnarray}
which cannot be written in the form $r=r(\theta)$ (see Gregory {\it et al.}, in prep. for a full derivation of this result).  Equation (\ref{dipoctshape}) reduces to the result for a dipole magnetic loop, given above, when $B_{\rm oct}=0$.  Parallel dipole-plus-octupole magnetic fields can also have higher latitude shells of closed field lines, see Figure \ref{fieldlineplot}. These magnetic field lines do not pass through the midplane and are described by a different equation that is not needed here as in our magnetic field geometry they do not carry accretion flow.     

Using equation (\ref{dipoctshape}) to determine points along a loop from the disk midplane at $r=r_m$ to the stellar surface at $r=R\ast$, we can calculate the field strength $B = (B_r^2+B_\theta^2)^{1/2}$ at any point along the loop using equations (\ref{equ_Br}) and (\ref{equ_Bt}).  In Figure \ref{Bfieldlines} we plot the field line shape for two different values of $B_{\rm oct}/B_{\rm dip}$, as well as the field strength along magnetic loops from the disk to the star.  In each plot, the dashed red line is a dipole magnetic loop that threads the disk at the same $r_m$ as the dipole-plus-octupole magnetic loop.  The influence of the octupole component on the field lines is clear, with their shape becoming more and more distorted from that of a dipole while approaching the star and for larger values of $B_{\rm oct}/B_{\rm dip}$.   

At the inner disk truncation radius $B$ is well approximated using the strength the dipole component alone (see section \ref{disk}). However, the influence of the octupole component is clear as we approach the star, where $B$ can significantly exceed that found for a dipole.  As we discuss in the following section, $B$ at the accretion shock can reach several kilo-Gauss, even for stars with sub-kilo-Gauss dipole components.


\section{B in the accretion shock}\label{shock}
First reported for an accreting PMS star by \citet{joh99}, strong circular polarisation can be measured in accretion-related emission lines, such as HeI 5876{\AA }.  This particular line, which forms in the accretion shock, probes the magnetic field where accreting gas impacts the star.  Independent studies have measured longitudinal fields of $\gtrsim$6\,${\rm kG}$ using the HeI 5876{\AA } emission line for the accreting PMS star GQ~Lup \citep{don12,joh13}, a star with a dipole component of only $\sim$1$\,{\rm kG}$. Longitudinal fields measured from the accretion hot spots are more typically $\sim$1-$\sim$4$\,{\rm kG}$ (e.g. \citealt{don10b,don11a,don13,che13}), at least for stars with large-scale magnetospheres well described by a dipole plus an octupole component.  Such field strengths are, commonly, well in excess of the polar strengths of the dipole components, ranging from $\sim$0.3 to $\sim$1.9$\,{\rm kG}$ for the same stars.  

Although the polar strength of the dipole component ($B_{\rm dip}$) provides a good estimate of $R_t$ in most cases, see section \ref{disk}, if it is assumed that the stellar magnetic field is a dipole, and therefore that the accreting field lines are dipolar, $B$ at the accretion shock can be severely underestimated. As measured from the accretion-related emission lines, $B$ where material impacts the star can be several kilo-Gauss, even for stars where the dipole component itself is only a few hundred Gauss. This is due to the presence of higher order magnetic field components. Likewise, if $B$ in the accretion hot spot is assumed to be representative of the dipole component, the disk truncation will be greatly overestimated.  

As one example, in Figure \ref{Bfieldlines} (left hand panels) $B_{\rm dip}=0.5\,{\rm kG}$ and $B_{\rm oct}=1.5\,{\rm kG}$.  At the stellar surface, for the considered field lines, $B$ at the accretion shock is $\sim$1.8$\,{\rm kG}$, while it is four times smaller, $\sim$450$\,{\rm G},$ if pure dipole magnetic field lines are used.  If we assume that the star has a dipole magnetic field and that 1.8$\,{\rm kG}$ at the accretion hot spot is representative of $B_{\rm dip}$, then the disk truncation radius would be overestimated [see equation (\ref{Rtdip})] by a factor of $4^{4/7}\approx 2.2$.\footnote{As the accretion spot is not at the pole, $B_{\rm dip}$ would be even larger, increasing the overestimation of $R_t$.}  



\section{Conclusions}\label{conclusions}
Models of accretion flow, of the star-disk interaction, and of accretion shocks should incorporate multipolar magnetic fields.  Dipole magnetic fields provide a poor representation of the true magnetic complexity of PMS stars.  Even AA Tau, whose magnetic field is closest to a dipole \citep{don10b}, has a non-negligible $\sim$0.5$\,{\rm kG}$ octupole component.  However, the large-scale magnetic fields of many accreting PMS stars are still somewhat simple, being dominantly axisymmetric and well-described by a (tilted) dipole component plus a (tilted) octupole component \citep{gre11}.  Some of the best studied PMS stars have such magnetic field topologies, including AA~Tau, BP~Tau, V2129~Oph, TW~Hya, and others, although other higher order magnetic modes, and non-axisymmetric components are present too.  Other stars, typically those that have developed large radiative cores, host more complex, multipolar, and non-axisymmetric large-scale magnetic fields \citep{hus09,gre12,gre14}.       

In this conference proceedings we used a simple model of a star with a dipole plus an octupole component.  In order to make progress analytically, we assumed that both magnetic moments were aligned with the stellar rotation axis, and were parallel (the main positive pole of the octupole coincident with the main positive pole of the dipole).\footnote{Some stars, such as AA~Tau and TW~Hya, have field configurations that are closer to an anti-parallel dipole-plus-octupole, where the main positive pole of the octupole is close to coincident with the main negative pole of the dipole.  For brevity we have not considered such magnetic fields in this work.  Details can be found in \citet{gre11} and Gregory {\it et al.}, in prep.}  Although these models are still simplified, they provide a far more realistic approximation to the true complexity of the magnetic fields of many accreting PMS stars than what can be achieved with a dipole.  We have shown that:

\begin{itemize}
\item In most cases, as the higher order magnetic components decay faster with distance from the stellar surface, the disk truncation radius can be well approximated by using the polar strength of the dipole component alone. However, there are exceptions, including: i) when the mass accretion rate is large; ii) when the dipole component is weak; iii) when the higher order magnetic field components are very strong; iv) when the large-scale magnetosphere of the star is highly multipolar or tilted; and v) some combination of all of these which will result in a smaller disk truncation radius, where the impact of higher order magnetic components is larger.
\item For the parallel dipole-plus-octupole magnetic fields, when the disk is truncated close to the magnetic null point, small changes in the mass accretion rate or the strengths of the magnetic field components can result in all of, or a portion of, the accretion flow impacting the star in low latitudes hot spots.  This diversion of material from high to low latitude hot spots will alter the stellar variability.
\item Although $B_{\rm dip}$ can often be used to calculate $R_t$, $B$ along the magnetic loops departs strongly from that of dipole magnetic field lines, as does the shape of the magnetic loops.
\item $B$ in the accretion shock can reach multiple kilo-Gauss, even for stars with dipole components of only a few hundred Gauss.
\item If the high field strengths measured in accretion hot spots are erroneously taken to be representative of the strength of the dipole component, and a dipole magnetic field is assumed, then the disk truncation radius will be overestimated. Likewise, use of the dipole component alone will often result in a significant underestimation of $B$ at the accretion shock.
\end{itemize}

In this work we have considered the impact of magnetic fields consisting of a dipole plus an octupole component on the disk truncation radius, $B$ along the accretion flow, and $B$ at the accretion shock.  Dropping the observationally unrealistic assumption that accreting PMS stars have dipole magnetic fields has several additional effects on magnetospheric accretion, the star-disk system, and the stellar rotational evolution, which we have not discussed here.  For example,   
the specific angular momentum transferred to the star through the star-disk interaction is an order of magnitude less for stars with octupole dominated fields compared to those with dominantly dipolar magnetic fields \citep{bat13}.  For multipolar magnetic fields, and including the dipole-plus-octupole magnetic fields considered here, material accretes into smaller hot spots, with a (usually) smaller accretion filling factor (e.g. \citealt{ada12}).  The accretion flow being funnelled into smaller spots increases the pre-shock density of the hot spots \citep{gre07,gre08,ada12} and increases their temperature \citep{ada12}.  Although we can use $B_{\rm dip}$ to calculate $R_t$ in most cases, the dipole component alone provides a poor representation of the structure of accretion flow, of $B$ along accretion columns, and of $B$ where material impacts the star. Models of magnetospheric accretion, of accretion flow, and of accretion shocks, must incorporate multipolar magnetic fields.


\section*{Acknowledgements}
{SGG acknowledges support from the Science \& Technology Facilities Council (STFC) via an Ernest Rutherford Fellowship [ST/J003255/1]. JFD and GAJH warmly thank the IDEX initiative at Universit{\'e} F{\'e}d{\'e}rale Toulouse Midi-Pyr{\'e}n{\'e}es (UFTMiP) for generous funding related to this research project.}

\bibliographystyle{cs19proc}
\bibliography{sgregory_magnetic.bib}

\end{document}